\documentclass[10pt, reqno]{amsart}

\usepackage[hmargin=0.8in,twosideshift=0in,height=8.6in]{geometry}
\usepackage{amssymb,amsthm}
\usepackage{delarray,verbatim}
\usepackage{natbib}

\usepackage{ifpdf}
\ifpdf
\usepackage[pdftex]{graphicx}
\else
\usepackage[dvips]{graphicx}
\fi

\usepackage{ifpdf}
\usepackage{color}
\definecolor{webgreen}{rgb}{0,.5,0}
\definecolor{webbrown}{rgb}{.8,0,0}
\definecolor{emphcolor}{rgb}{0.95,0.95,0.95}

\usepackage{hyperref}
\hypersetup{%
          colorlinks=true,
          linkcolor=webbrown,
          filecolor=webbrown,
          citecolor=webgreen,
          breaklinks=true}
\ifpdf
\hypersetup{pdftex,
            pdfstartview=FitH, 
            bookmarksopen=true,
            bookmarksnumbered=true
}
\else
\hypersetup{dvips}
\fi

\usepackage{multicol}
\usepackage{multirow}

\numberwithin{equation}{section} 

\newtheorem{remark}{Remark}[section]
\newtheorem{proposition}{Proposition}[section]

\newcommand{\E}{\mathbb{E}}

\newcommand{\R}{\mathbb{R}}
\renewcommand{\S}{\mathcal{S}}

\renewcommand{\L}{\mathcal{L}}
\newcommand{\A}{\mathcal{A}}

\title[On the Pricing of American Options for Jump Diffusions]{Pricing American Options for Jump Diffusions by Iterating Optimal Stopping Problems for Diffusions}

\author{Erhan Bayraktar}
\thanks{Department of Mathematics, University of Michigan, Ann Arbor, MI 48109, USA; e-mail:\{erhan,haoxing\}@umich.edu}
\thanks{This research  is supported in part by the National Science Foundation. }

\author{Hao Xing }


\begin{document}

\begin{abstract}
We approximate the price of the American put for jump diffusions by a sequence of functions, which are computed iteratively. This sequence converges to the price function uniformly and exponentially fast. Each element of the approximating sequence solves an optimal stopping problem for geometric Brownian motion, and can be numerically computed using the classical finite difference methods. We prove the convergence of this numerical scheme and present examples to illustrate its performance.

\end{abstract}

\maketitle
\section{Introduction}
Jump diffusion models are heavily used in modeling stock prices since they can capture the excess kurtosis and skewness of the stock price returns, and  they can produce the smile in the implied volatility curve (see \cite{cont-tankov}).  Two well-known examples of these models are i) the model of \cite{merton}, in which the jump sizes are log-normally distributed, and ii) the model of \cite{kou-wang}, in which the logarithm of jump sizes have the so called double exponential distribution.  Based on the results of \cite{bayraktar-fin-horizon} we propose a numerical algorithm to calculate the American option prices for jump diffusion models
and analyze the convergence behavior of this algorithm.

As observed by \cite{bayraktar-fin-horizon}, we can construct an increasing sequence of functions, which are value functions  of optimal stopping problems (see \eqref{eq:defn-v-n} and also \eqref{eq:alt}), that converge to the price function of the American put option uniformly and exponentially fast. Because each element of this sequence solves an optimal stopping problem it shares the same regularity properties, such as convexity and smoothness, with the original price function. Even the corresponding free boundaries have the same smoothness properties (when they have a discontinuity, which can only happen at maturity, the magnitude of the discontinuity is the same; see \eqref{eq:asy-s}).
Therefore, the elements in this approximating sequence provide a good imitation to the value function besides being close to it numerically (see Remark~\ref{rem:imptrem}).
On the other hand,
each of these functions can be represented as classical solutions of free boundary problems (see \eqref{eq:pde-v-n}) for geometric Brownian motion, and therefore can be implemented using classical finite difference methods.  We build an iterative numerical algorithm based on discretizing these free boundary problems (see \eqref{eq:diffeq-un}). When the mesh sizes are fixed, we show that the iterative sequence we constructed is monotonous and converges uniformly and exponentially fast (see Proposition~\ref{prop:exp-conv}). We also show, in a rather direct way, that when the mesh sizes go to zero our algorithm converges to the true price function (see Proposition~\ref{prop:u-tildeu}).

The pricing in the context of jump models is difficult since the
prices of  options satisfy integro-partial differential equations
(integro-pdes), i.e. they have non-local integral terms, and the
usual finite-difference methods are not directly applicable because the integral term leads to full matrices.
Recently there has been a lot of interest in developing numerical
algorithms for pricing in jump models, see e.g.  \cite{ait-run},
\cite{A-O}, \cite{and-and}, \cite{cont-volt},   \cite{dFL},
\cite{hir-madan},  \cite{jaimungal08}, \cite{kpw},
\cite{kou-wang}, \cite{metwally}, \cite{zhang}, among them
\cite{A-O}, \cite{cont-volt}, \cite{hir-madan} and
\cite{jaimungal08} treated specific or general jump models with
infinite activity jumps. These algorithms have been extensively discussed in Chapter 12 of \cite{cont-tankov}. In this paper, relying on the results of \cite{bayraktar-fin-horizon} as described above, we give an efficient numerical algorithm (and analyze its error versus accuracy characteristics) to efficiently compute American option prices for jump diffusion models with finite activity. One can handle infinite activity models by increasing the volatility coefficient appropriately as suggested on p. 417 of \cite{cont-tankov}.

An ideal numerical algorithm, which is
most often an iterative scheme, *should monotonically converge to
the true price uniformly (across time and space) and exponentially
fast*, that is, the error bounds should be very tight. This is the
only way one can be sure that the price output of the algorithm is
close to the true price after a reasonable amount of runtime and
without having to compare the price obtained from the algorithm to
other algorithms' output.
 It is also desirable to obtain a scheme **that does not deviate from the numerical pricing schemes, such as finite difference methods, that were developed for models that do not account for jumps**. Financial engineers working in the industry are already familiar with finite difference schemes such as projected successive over relaxation, PSOR, (see e.g. \cite{dewyne}) and Brennan-Schwartz algorithm (see \cite{brennan-schwartz} and \cite{j-l-l}) to solve the partial differential equations associated with free boundary problems, but may not be familiar with the intricacies involved in solving integro-partial differential equations developed in the literature. It would be ideal for them if they could use what they already know with only a slight modification to solve for the prices in a jump diffusion model. In this paper, we develop an algorithm which establishes both * and **. In Section  \ref{sec:numericalper}, we will name this algorithm, depending on which classical method we
 use to solve the sparse linear systems in \eqref{eq:diffeq-un}, as either ``Iterated PSOR" or ``Iterated Brennan-Schwartz".

In the next section we introduce a sequence of optimal stopping problems that approximate the price function of the American options, and discuss their properties.
 In Section 3, we introduce a numerical algorithm and analyze its convergence properties. In the last section we
give numerical examples to illustrate the competitiveness of our algorithm and price
American, Barrier and European options for the models of
\cite{kou-wang} and \cite{merton}.

\section{A Sequence of Optimal Stopping Problems for Geometric Brownian Motion Approximating the American Option Price For Jump Diffusions}
\label{sec:iterated-PSOR}

We will consider a jump diffusion model for the stock price $S_t$ with $S_0=S$, and assume that return process $X_t:=\log(S_t/S)$, under the risk neutral measure, is given by
\begin{equation}\label{eq:dyn}
dX_t=\left(\mu-\frac{1}{2}\sigma^2\right) dt+\sigma dW_t+
\sum_{i=1}^{N_t} Z_i, \quad X_0=0.
\end{equation}
In \eqref{eq:dyn}, $\mu=r+\lambda-\lambda \xi$, $r$ is the
risk-free rate, $W_t$ is a Brownian motion, $N_t$ is a
Poisson process with rate $\lambda$ independent of the Brownian
motion, $Z_i$ are independent and identically distributed, and
come from a common distribution $F$ on $\R$, that satisfies $\xi
:=\int_{\mathbb{R}}e^{z}F(dz)<\infty$. The last condition guarantees that the
stock prices have finite expectation. We will assume that the volatility
$\sigma$ is strictly positive. The price function of the
American put with strike price $K$ is
\begin{equation}\label{eq:opt-stop}
V(S,t):=\sup_{\tau \in \S_{t,T}}\E\{e^{-r
(\tau-t)}(K-S_{\tau})^+\big|S_t=S\},
\end{equation}
in which $\S_{t,T}$ is the set of stopping times of the filtration generated by $X$ that belong to the interval $[t,T]$ ($t$ it the current time, $T$ is the maturity of the option).  Instead of working with the pricing function $V$ directly, which is the
unique classical solution of the following  integro-differential free boundary problem (see Theorem 3.1 of \cite{bayraktar-fin-horizon})
\begin{equation}\label{eq:pde-v}
\begin{split}
& \frac{\partial}{\partial t} V(S,t)+\A V (S,t) +\lambda \cdot
\int_{\R}V(e^{z}\cdot S,t)F(dz)-(r+\lambda)\cdot V (S,t)  =0 \quad
S> s(t),
\\ & V(S,t) =K-S, \quad S \leq s(t), \\
& V(S,T) = \left(K-S\right)^+,
\end{split}
\end{equation}
in which, $\mathcal{A}$ is the differential operator
\begin{equation}\label{eq:inifinite}
\mathcal{A}:=\frac{1}{2}\sigma^2 S^2 \frac{d^2}{dS^2}+\mu S
\frac{d}{dS},
\end{equation}
 and $t \rightarrow s(t)$, $t \in [0,T]$, is the exercise boundary that needs to be determined along with the pricing function $V$;
we will construct a sequence of pricing problems for the geometric Brownian motion
\begin{equation}
dS^0_t=\mu S^0_t dt+\sigma S_t^0 dW_t, \quad S^0_0=S.
\end{equation}
To this end, let us introduce a functional operator $J$, whose action on a test
function $f:\R_+ \times [0,T]   \rightarrow \R_+$
is the solution of the following pricing problem for the geometric Brownian motion: $(S^0_t)_{t \geq 0}$
\begin{equation}\label{eq:defn-Jf}
J f (S,t)=\sup_{\tau \in
\tilde{\S}_{t,T}}\E\left\{\int_t^{\tau}e^{-(r+\lambda)(u-t)}\lambda \cdot Pf
(S_u^0,u)du+
e^{-(r+\lambda)(\tau-t)}(K-S^0_{\tau})^+\big|S^0_t=S\right\},
\end{equation}
in which
\begin{equation}\label{eq:defn-Sf}
Pf(S,u)=\int_{\R}f(e^{z}\cdot S,u) F(dz)=\E[f(e^{Z} S,u)], \quad S\geq 0,
\end{equation}
for a random variable $Z$ whose distribution is $F$, and $\tilde{\S}_{t,T}$ is the set of stopping times of the filtration generated by $W$ that take values in $[t,T]$. Let us define a sequence of pricing functions by
\begin{equation}\label{eq:defn-v-n}
v_0(S,t)=(K-S)^+, \quad v_{n+1}(S,t)=J v_{n}(S,t),\; n \geq 0,
\quad \text{for all $(S,t) \in \R_+ \times [0,T]$.}
\end{equation}
For each $n \geq 1$, the pricing function $v_n$ is the unique solution of the classical free-boundary problem (instead of a free boundary problem with an integro-diffential equation)
\begin{equation}\label{eq:pde-v-n}
\begin{split}
& \frac{\partial}{\partial t} v_{n}(S,t)+\A v_{n}(S,t)
-(r+\lambda)\cdot v_{n} (S,t)  =-\lambda \cdot (P v_{n-1})(S,t),
\quad S> s_{n}(t),
\\ & v_{n}(S,t) = K-S, \quad S \leq s_{n}(t),\\
& v_n(S,T) = (K-S)^+,
\end{split}
\end{equation}
in which $t \rightarrow s_n(t)$ is the free-boundary (the optimal
exercise boundary) which needs to be determined (see Lemma 3.5 of
\cite{bayraktar-fin-horizon}). Now starting from $v_0$, we can
calculate  $\{v_n\}_{n\geq 0}$  sequentially. For $v_n$, the
solution of (\ref{eq:pde-v-n}) can be determined using a classical
finite difference method (we use the Crank-Nicolson discretization
along with Bernnan-Schwartz algorithm or PSOR in the the following
sections) given that the function $v_{n-1}$ is available. The term
on the right-hand-side of (\ref{eq:pde-v-n}) can be computed
either using Monte-Carlo or a numerical integrator (we use the
numerical integration with the Fast Fourier Transformation (FFT)
in our examples). Iterating the solution for \eqref{eq:pde-v-n} a
few times we are able to obtain the American option price $V$
accurately since the sequence of functions $\{v_n\}_{n \geq 0}$
converges to $V$ uniformly and exponentially fast:
\begin{equation}\label{eq:iterror}
v_{n}(S,t) \leq V(S,t) \leq v_{n}(S,t)+K
\left(1-e^{-(r+\lambda)(T-t)}\right)^n\left(\frac{\lambda}{\lambda+r}\right)^n
, \quad S \in \R_+,\; t \in (0,T),
\end{equation}
see Remark 3.3 of \cite{bayraktar-fin-horizon}. Note that the usual values of $T$ for the traded options is $0.25$, $0.5$, $0.75$, $1$ year.

\begin{remark}\label{rem:imptrem}The approximating sequence $\{v_n\}_{n \geq 0}$ goes beyond  approximating the value function $V$.
Each $v_n$ and its corresponding free boundary have the same
regularity properties which $V$ and its corresponding free
boundary have. In a sense, for large enough $n$, $v_n$ provides a
good imitation of $V$. Below we list these properties:
\begin{enumerate}
\item[1)] The function $v_n$ can be written as the value function of an optimal stopping problem:
\begin{equation}\label{eq:alt}
v_n(S,t):=\sup_{\tau \in \S_{t,T}}\E\{e^{-r
(\tau \wedge \sigma_n-t)}(K-S_{\tau \wedge \sigma_n})^+\big|S_t=S\},
\end{equation}
in which $\sigma_n$ is the n-th jump time of the Poisson process
$N_t$. \item[2)] Each $v_n$ is a convex function in the
$S$-variable, which is a property that is also shared by $V$.
 Moreover, the sequence $\{v_n\}_{n\geq 0}$ is a monotone increasing sequence converging to the value function $V$ (see \eqref{eq:iterror}). \\
\item[3)] The free boundaries $s(t)$ and $s_n(t)$ have the same
regularity properties (see \cite{bayraktar-xing} and its
references):
\begin{enumerate}
\item[a)] They are strictly decreasing. \item[b)] They may exhibit
discontinuity at $T$: If the parameters satisfy
\begin{equation}\label{eq:condition-c}
r< \lambda \int_{\R_+} \left(e^z-1\right)F(dz),
\end{equation}
we have
\begin{equation}\label{eq:asy-s}
 \lim_{t\rightarrow T} s(t) = \lim_{t\rightarrow T} s_n(t)= S^*<K,
 \quad n\geq 1,
\end{equation}
where $S^*$ is the unique solution of the following integral
equation
\begin{equation} \label{eq:equation-s*}
 - rK +\lambda \int_{\R} \left[\left(K-Se^z\right)^+ -
 \left(K-Se^z\right)\right]F(dz) = 0.
\end{equation}
 We will see such an example in Section \ref{sec:numericalper},
 where the equation (\ref{eq:equation-s*}) can be solved
 analytically for some jump distribution $F$.
 \item[c)] Both $s(t)$ and $s_n(t)$ are continuously differentiable on $[0,T)$.
 \end{enumerate}
\end{enumerate}

\end{remark}

\section{A Numerical Algorithm and its convergence analysis}

\subsection{The numerical algorithm}
In this section, we will discretize the algorithm introduced in
the last section and give more details. For the convenience of the
numerical calculation, we will first change the variable:
$x\triangleq\log{S}$, $x(t)\triangleq\log{s(t)}$ and
$u(x,t)\triangleq V(S,t)$. $u$ satisfies the following
integro-differential free boundary problem
\begin{equation}\label{eq:problem-u}
\begin{split}
 & \frac{\partial}{\partial t} u + \frac12 \sigma^2
 \frac{\partial^2}{\partial x^2} u +
 \left(\mu-\frac12\sigma^2\right)\frac{\partial}{\partial x} u
 -(r+\lambda)u + \lambda \cdot (Iu)(x,t) =0, \quad x>x(t)\\
 & u(x,t) = K-e^x, \quad x\leq x(t)\\
 &u(x,T)=(K-e^x)^+,
\end{split}
\end{equation}
in which
\begin{equation}\label{eq:def_I}
(Iu)(x,t) = \int_{\R} u(x+z,t)\rho(z) dz,
\end{equation}
with $\rho(z)$ as the density of the distribution $F$. Similarly,
$u_n(x,t)\triangleq v_n(S,t)$ satisfies the similar free boundary
problem where $u$ in (\ref{eq:problem-u}) is replaced by $u_n$ in
differential parts and by $u_{n-1}$ in the integral part. In
addition, it was shown in Theorem 4.2 of \cite{yang} that the free
boundary problem (\ref{eq:problem-u}) is equivalent to the
following variational inequality
\begin{equation}\label{eq:varineq-u}
\begin{split}
 & \L_{D} u(x,t) + \lambda \cdot (Iu)(x,t) \leq 0\\
 & u(x,t) \geq g(x) \\
 & \left[\L_{D} u(x,t) + \lambda \cdot
 (Iu)(x,t)\right]\cdot\left[u(x,t)-g(x)\right] = 0, \quad (x,t)\in
 \R\times [0,T],
\end{split}
\end{equation}
in which
\begin{equation*}
\begin{split}
 & \L_{D} u \triangleq \frac{\partial}{\partial t} u + \frac12 \sigma^2
 \frac{\partial^2}{\partial x^2} u +
 \left(\mu-\frac12\sigma^2\right)\frac{\partial}{\partial x} u
 -(r+\lambda)u \\
 & g(x) = \left(K-e^x\right)^+.
\end{split}
\end{equation*}
Since the second spacial derivative of $u$ does not exist along
the free boundary $x(t)$, the variational inequality
\eqref{eq:varineq-u} does not have a classical solution.
However, Theorem 3.2 of \cite{yang} showed that $u$ is the
solution of \eqref{eq:varineq-u} in the Sobolev sense. In the same
sense, $u_n(x,t)$ satisfies a similar variational inequality
\begin{equation}\label{eq:varineq-un}
\begin{split}
 & \L_{D} u_n(x,t) + \lambda \cdot (Iu_{n-1})(x,t) \leq 0\\
 & u_n(x,t) \geq g(x) \\
 & \left[\L_{D} u_n(x,t) + \lambda \cdot
 (Iu_{n-1})(x,t)\right]\cdot\left[u_n(x,t)-g(x)\right] = 0, \quad (x,t)\in
 \R\times [0,T].
\end{split}
\end{equation}

Let us discretize (\ref{eq:varineq-u}) using Crank-Nicolson
scheme. For fixed $\Delta t$, $\Delta x$, $x_{min}$ and $x_{max}$,
let $M\Delta t=T$ and $L\Delta x = x_{max}-x_{min}$. Let us denote
$x_l=x_{min}+l\Delta x$, $l=0, \cdots, L$. By $\tilde{u}^{l,m}$ we
will denote the solution of the following difference equation
\begin{equation}\label{eq:diffeq-u}
\begin{split}
 & -\theta p_- \tilde{u}^{l-1,m} + (1+\theta p_0) \tilde{u}^{l,m} -
 \theta p_+ \tilde{u}^{l+1,m} - \tilde{b}^{l,m} \geq 0 \\
 & \tilde{u}^{l,m} \geq g^l \\
 & \left[-\theta p_- \tilde{u}^{l-1,m} + (1+\theta p_0) \tilde{u}^{l,m} -
 \theta p_+ \tilde{u}^{l+1,m} - \tilde{b}^{l,m}\right] \cdot
 \left[\tilde{u}^{l,m} - g^l\right]=0,
\end{split}
\end{equation}
for $m=M-1, \cdots, 0$, $l=0, \cdots, L$, satisfying the terminal
condition $\tilde{u}^{l,M} = g^l= \left(K-e^{x_l}\right)^+$ and
Dirichlet boundary conditions. $\theta$ is the weight factor. When
$\theta =1$, the scheme (\ref{eq:diffeq-u}) is the completely
implicit Euler scheme; when $\theta =1/2$, it is the classical
Crank-Nicolson scheme. The coefficients $p_-$, $p_+$ and $p_0$ are
given by
\begin{equation}\label{eq:coeff-p}
\begin{split}
 & p_- = \frac12 \sigma^2 \frac{\Delta t}{(\Delta x)^2} - \frac12
 \left(\mu-\frac12 \sigma^2\right)\frac{\Delta t}{\Delta x},\\
 & p_+ = \frac12 \sigma^2 \frac{\Delta t}{(\Delta x)^2} + \frac12
 \left(\mu-\frac12 \sigma^2\right)\frac{\Delta t}{\Delta x},\\
 & p_0 =p_-+p_+ +(r+\lambda) \Delta t.
\end{split}
\end{equation}
The term $\tilde{b}$ is defined by
\begin{equation}\label{eq:def-b}
\tilde{b}^{l,m} = (1-\theta)p_- \tilde{u}^{l-1,m+1} +
(1-(1-\theta)p_0)\tilde{u}^{l,m+1} + (1-\theta)p_+
\tilde{u}^{l+1,m+1} + \lambda \Delta t\,\cdot
\left[(1-\theta)(\tilde{I}\tilde{u})^{l,m+1} +
\theta(\tilde{I}\tilde{u})^{l,m}\right].
\end{equation}
$\tilde{I}$ in (\ref{eq:def-b}) is the discrete version of the
convolution operator $I$ in (\ref{eq:def_I}). It will be
convenient to approximate this convolution integral using Fast
Fourier Transformation (FFT). Discretizing a sufficiently large
interval $[z_{min}, z_{max}]$ into $J$ sub-intervals. For the
convenience of the FFT, we will choose these $J$ sub-intervals
equally spaced, such that $J\Delta z = z_{max} - z_{min}$. We also
choose $\Delta x = \alpha \Delta z$, where $\alpha$ is a positive
integer, so that the numerical integral may have finer grid than
the grid in $x$. Let $z_{j} = z_{min} + j\Delta z$, $j=0, \cdots,
J$. $\tilde{I}$ is defined by
\begin{equation}\label{eq:def-tilde-I}
\left(\tilde{I}\tilde{u}\right)^{l,m} = \sum_{j=0}^{J-1}
\tilde{u}_{interp}\left(x_{l}+ z_{j},m\Delta
t\right)\rho(z_j)\Delta z,
\end{equation}
in which the value of $\tilde{u}_{interp}$ is determined by the
linear interpolation $\tilde{u}$. That is if there is some $l'$
satisfying
\[
 x_{l'} \leq x_{l}+ z_j \leq x_{l'+1},
\]
then
\[
 \tilde{u}_{interp}\left(x_{l}+z_j, m\Delta t\right) =
 (1-w)\tilde{u}^{l',m} + w \tilde{u}^{l'+1,m},
\]
for some $w\in[0,1]$. On the other hand, if $x_{l}+z_j$ is outside
the interval $[x_{min},x_{max}]$, the value of
$\tilde{u}_{interp}$ is determined by the boundary conditions.
Moreover, in (\ref{eq:def-tilde-I}) we also assume
\begin{equation}\label{eq:ass-rho}
\rho(z_{j}) \geq 0, \quad \text{ for all } j, \quad \text{ and }
\quad \sum_{j=0}^{J-1}\rho(z_j) \leq 1.
\end{equation}
Now (\ref{eq:def-tilde-I}) can be calculated using FFT. See
Section 6.1 in \cite{A-O} for implementation details.

Note that numerically solving the system (\ref{eq:diffeq-u}) is
difficult due to the contribution of the integral term
$\tilde{I}\tilde{u}$. Therefore, following the results in Section
\ref{sec:iterated-PSOR}, we will discretize (\ref{eq:varineq-un})
recursively (using the Crank-Nicoslon scheme) to obtain the
sequence $\{\tilde{u}_n\}_{n\geq 0}$ recursively. Let
$\tilde{u}^{l,m}_0 = g^{l}$. For $n \geq 1$, $\tilde{u}_n$ is
defined recursively by

\begin{equation}\label{eq:diffeq-un}
\begin{split}
 & -\theta p_- \tilde{u}_n^{l-1,m} + (1+\theta p_0) \tilde{u}_n^{l,m} -
 \theta p_+ \tilde{u}_n^{l+1,m} - \tilde{b}_n^{l,m} \geq 0 \\
 & \tilde{u}_n^{l,m} \geq g^l \\
 & \left[-\theta p_- \tilde{u}_n^{l-1,m} + (1+\theta p_0) \tilde{u}_n^{l,m} -
 \theta p_+ \tilde{u}_n^{l+1,m} - \tilde{b}_n^{l,m}\right] \cdot
 \left[\tilde{u}_n^{l,m} - g^l\right]=0,
\end{split}
\end{equation}
with the terminal condition $\tilde{u}_n^{l,M}= g^l$ and Dirichlet
boundary conditions. Similar to (\ref{eq:def-b}), $\tilde{b}_n$ is
defined by
\begin{equation}\label{eq:def-bn}
\begin{split}
\tilde{b}_n^{l,m} = & (1-\theta)p_- \tilde{u}_n^{l-1,m+1} +
(1-(1-\theta)p_0)\tilde{u}_n^{l,m+1} + (1-\theta)p_+
\tilde{u}_n^{l+1,m+1}\\
& + \lambda \Delta t\cdot
\left[(1-\theta)(\tilde{I}\tilde{u}_{n-1})^{l,m+1} +
\theta(\tilde{I}\tilde{u}_{n-1})^{l,m}\right].
\end{split}
\end{equation}
For each $n$, we will solve the sparse linear system of equations
(\ref{eq:diffeq-un}) using the projected PSOR method (see eg. \cite{dewyne}).

\begin{remark}\label{rem:locglb}
We will iterate (\ref{eq:diffeq-un}) to approximate the solution
of \eqref{eq:diffeq-u}, which can be seen as a global fixed point
iteration algorithm. This global fixed point algorithm is
different from the local fixed point algorithm in \cite{dFL},
where d'Halluin et al. implemented the Crank Nicolson time
stepping of a non-linear integro-partial differential equation
coming from an alternative representation (due to the penalty
method) of the American option price function. Also see \cite{dFV}
for the case of European options. Note that discretizing the
non-linear PDE that arises from the penalized formulation
introduces an extra error. We work with the variational
formulation directly.

Each $\tilde{u}_n$ approximates $u_n$, which itself is the value
function of an optimal stopping problem, and as we have discussed
in Remark~\ref{rem:imptrem}  provides a good imitation of the
American option price function. Each of these iterations provide
strictly decreasing free boundary curves with the same regularity
and jump properties as the free boundary curve for the American
option price function, see Remarks \ref{rem:imptrem} and
\ref{rem:lstrm}. The approximating sequence in \cite{dFL} does not
carry the same meaning, it is a technical step to carry out the
Crank Nicolson time stepping of their non-linear integro-PDE.

\end{remark}

\subsection{Convergence of the Numerical Algorithm}

In the following, we will show the convergence of the numerical
algorithm for the completely implicit Euler scheme ($\theta=1$).
We first show that $\{\tilde{u}_n\}_{n\geq 0}$ is a monotone
increasing sequence. Extra care has to be given to make the approximating sequence monotone in the penalty formulation of \cite{dFL} (see Remark 4.3 on page 341), but the monotonicity comes out naturally in our formulation. Next, we prove that the sequence
$\{\tilde{u}_n\}_{n\geq 0}$ is uniformly bounded above by the
strike price $K$ and converges to $\tilde{u}$ at an exponential
rate. At last, we will argue that as the mesh sizes $\Delta x$ and
$\Delta t$ go to zero $\tilde{u}$ converges to the American
option value function $u$. In the following four propositions, we
let  $\Delta t$ and $\Delta x$ to be sufficiently small so that constants $p_-$ and $p_+$ defined in (\ref{eq:coeff-p}) are
positive.

\begin{proposition}\label{prop:u-n-increase}
 The sequence $\{\tilde{u}_n\}_{n\geq 0}$ is a monotone increasing sequence.
\end{proposition}
\begin{proof}
 When $\theta =1$, subtracting the third equality for n-th iteration in
 (\ref{eq:diffeq-un})  from the equality for
 $(n+1)$-th iteration, we obtain
 \begin{equation}\label{eq:u_n+1 - u_n}
 \begin{split}
  & \left[-p_- \tilde{u}_n^{l-1,m} + (1+p_0) \tilde{u}_n^{l,m} - p_+\tilde{u}_n^{l+1,m} -
  \tilde{b}_n^{l,m}\right] \left[\tilde{u}_{n+1}^{l,m} -
  \tilde{u}_n^{l,m}\right] \\
  & + \left\{-p_- \left(\tilde{u}_{n+1}^{l-1,m} - \tilde{u}_n^{l-1,m}\right) + \left(1+p_0\right)\left(\tilde{u}_{n+1}^{l,m}-\tilde{u}_n^{l,m} \right) - p_+ \left(\tilde{u}_{n+1}^{l+1,m} -
  \tilde{u}_n^{l+1,m}\right)\right.\\
  & - \left. \left(\tilde{u}_{n+1}^{l,m+1} - \tilde{u}_n^{l,m+1}\right) - \lambda\Delta t \cdot \left(\tilde{I}\left(\tilde{u}_n - \tilde{u}_{n-1}\right)\right)^{l,m}
  \right\}
  \left[\tilde{u}^{l,m}_{n+1}- g^l\right]=0.
 \end{split}
 \end{equation}
 in which we used the linearity of the operator $\tilde{I}$.
 Let us define the vectors
 \begin{eqnarray*}
 e^m_{n+1} &=& \left(\tilde{u}_{n+1}^{0,m} -
\tilde{u}_{n}^{0,m}, \cdots, \tilde{u}_{n+1}^{L,m} -
 \tilde{u}_n^{L,m}\right)^T,\\
 f^m_{n+1} &=& \left(\left[(\tilde{u}_{n+1}^{0,m+1} - \tilde{u}_n^{0,m+1}) + \lambda \Delta t\, \cdot \left(\tilde{I} (\tilde{u}_n -\tilde{u}_{n-1})\right)^{0,m}\right]\left[\tilde{u}^{0,m}_{n+1} - g^0\right],
 \cdots, \right.\\
 & & \left. \left[(\tilde{u}_{n+1}^{L,m+1} - \tilde{u}_n^{L,m+1}) + \lambda \Delta t\, \cdot \left(\tilde{I} (\tilde{u}_n -\tilde{u}_{n-1})\right)^{L,m}\right]\left[\tilde{u}^{L,m}_{n+1} -
 g^L\right]\right)^T
  .
 \end{eqnarray*}
Equation (\ref{eq:u_n+1 - u_n}) can be represented as
 \begin{equation}\label{eq:u_n+1 - u_n mat}
 A \, e^m_{n+1} = f^m_{n+1},
 \end{equation}
 in which the matrix $A$'s entries are
 \begin{equation*}
 \begin{split}
  a_{l,j} = \left\{
  \begin{array}{cl}
   -p_- \left(\tilde{u}^{l,m}_{n+1} - g^{l} \right) & j= l-1 \\
   (1+p_0)\left(\tilde{u}^{l,m}_{n+1}-g^l\right) + \left(-p_- \tilde{u}^{l-1, m}_{n} + (1+p_0)\tilde{u}^{l,m}_{n} - p_+ \tilde{u}^{l+1,m}_{n} -
   \tilde{b}^{l,m}_n\right) & j=l \\
   -p_+ \left(\tilde{u}_{n+1}^{l,m}-g^l\right) & j=l+1 \\
   0 & \text{others}.
  \end{array}
  \right.
 \end{split}
 \end{equation*}
 On the other hand, using the first and second inequalities in
 (\ref{eq:diffeq-un}) and the fact that $p_-$ and $p_+$ are positive, we see that $A$ is an M-matrix, i.e. $A$ has positive
 diagonals, non-positive off-diagonals and the row sums are positive.
 As a result all entries of $A^{-1}$ are nonnegative.

 Now we can prove the proposition by induction. Note that $\tilde{u}_1 \geq \tilde{u}_0 =
 g$, as a result of the second inequality in (\ref{eq:diffeq-un})
 and the definition of $\tilde{u}_0$. Assuming $\tilde{u}_n\geq
 \tilde{u}_{n-1}$, we will show that $\tilde{u}_{n+1}\geq \tilde{u}_n$, i.e. $\tilde{u}^{l,m}_{n+1} - \tilde{u}^{l,m}_n \geq
 0$ for all $l$ and $m$, in the following.

 First, the terminal condition of $\tilde{u}_n$ gives us $\tilde{u}_{n+1}^{l,M} -
 \tilde{u}_{n}^{l,M}=0$. Second, $\left(\tilde{I}(\tilde{u}_n -
 \tilde{u}_{n-1})\right)^{l,m}$ is nonnegative from the assumption
 (\ref{eq:ass-rho}). Assuming $\tilde{u}_{n+1}^{l,m+1} -
 \tilde{u}_n^{l,m+1}$ nonnegative, we have $f^m_{n+1}$ in (\ref{eq:u_n+1 - u_n
 mat}) as a nonnegative vector. Combining with the fact that all
 entries of $A^{-1}$ are nonnegative, the nonnegativity of
 $\tilde{u}_{n+1}^{l,m}-\tilde{u}_{n}^{l,m}$ follows from
 multiplying $A^{-1}$ on both sides of (\ref{eq:u_n+1 - u_n mat}).
 Then the result follows from an induction $m$.
\end{proof}

\begin{proposition}\label{prop:uniform-bound}
 $\{\tilde{u}_n\}_{n\geq 0}$ are uniformly bounded above by the strike price K.
\end{proposition}
\begin{proof}
 When $\theta =1$, in the third equality of
 (\ref{eq:diffeq-un}), there are some
 $(l,m)$ such that $\tilde{u}^{l,m}_n = g^l$. Otherwise we have
 \[
  (1+p_0) \tilde{u}^{l,m}_n = p_- \tilde{u}^{l-1,m}_n + p_+
  \tilde{u}^{l+1,m}_n + \tilde{u}^{l,m+1}_n + \lambda \Delta t
  \left(\tilde{I}\tilde{u}_{n-1}\right)^{l,m}.
 \]
However, in both cases, we obtain the following inequality
\begin{equation}\label{eq:un-bounded}
 (1+p_0) \left|\tilde{u}_n^{l,m}\right| \leq p_- B_n^m + p_+ B_n^m
 + B_n^{m+1} + \lambda \Delta t B_{n-1} + r\Delta t K, \quad 0\leq
 l\leq L, 0\leq m\leq M-1,
\end{equation}
in which we define
\[
 B_n^m = \left(\max_l{\left|\tilde{u}^{l,m}_n\right|}\right)
 \bigvee K, \quad B_n = \max_m{B_n^m}.
\]
Note that the right hand side of (\ref{eq:un-bounded}) is
independent of $l$. Moreover, $(1+p_0)K$ is also less than or
equal to the right hand side of (\ref{eq:un-bounded}). Therefore,
(\ref{eq:un-bounded}) gives us
\begin{equation}\label{eq:un-bounded-2}
 \left(1+(r+\lambda)\Delta t\right) B_n^m \leq B_n^{m+1} + \lambda
 \Delta t B_{n-1} + r\Delta t K.
\end{equation}
Given $B_n^{m+1} \leq K$ and $B_{n-1}\leq K$, it clear from
(\ref{eq:un-bounded-2}) that $B_n^m \leq K$. Now the proposition
follows from double induction on $m$ and $n$ with initial steps
$\tilde{u}_n^M = g \leq K$ and $\tilde{u}_0= g \leq K$.
\end{proof}

As a result of Propositions \ref{prop:u-n-increase}, we can define
\begin{equation}\label{eq:def-tu-inf}
 \tilde{u}_{\infty}^{l,m} = \lim_{n\rightarrow +\infty}
 \tilde{u}_n^{l,m}, \quad 0\leq l\leq L, 0\leq m\leq M.
\end{equation}
It follows from Proposition \ref{prop:uniform-bound} that
$\tilde{u}^{l,m}_{\infty}\leq K$. Letting $n$ go to $+\infty$, we can see
from (\ref{eq:diffeq-un}) that $\tilde{u}_{\infty}$ satisfies the
difference equation (\ref{eq:diffeq-u}). Therefore,
\begin{equation}\label{eq:u-infty-u}
 \tilde{u}_{\infty} = \tilde{u}.
\end{equation}
In the following, we will study the convergence rate of
\{$\tilde{u}_n\}_{n\geq 0}$.

\begin{proposition}\label{prop:exp-conv}
 $\tilde{u}_n$ converges to $\tilde{u}$ uniformly and
\begin{equation}\label{eq:error-conv}
 \max_{l,m}{\left(\tilde{u}^{l,m} - \tilde{u}^{l,m}_n\right)} \leq \left(1-\eta^M\right)^n
  \left(\frac{\lambda}{\lambda+r}\right)^n \tilde{K},
\end{equation}
where $\eta = \frac{1}{1+(\lambda+r)\Delta t} \in (0,1)$,
$\tilde{K}$ is a positive constant.
\end{proposition}
\begin{proof}
 Let us define
 \[
  e^{l,m}_n = \tilde{u}^{l,m} - \tilde{u}^{l,m}_n, \quad E^m_n =
  \max_l e^{l,m}_n, \quad E_n= \max_m E^m_n.
 \]
 Proposition \ref{prop:u-n-increase} and (\ref{eq:u-infty-u})
 ensure that $e^{l,m}_n$ is nonnegative. Moreover $e_n$ satisfies
 \begin{equation}\label{eq:u - u_n}
 \begin{split}
  & \left[-p_- \tilde{u}_n^{l-1,m} + (1+p_0) \tilde{u}_n^{l,m} - p_+\tilde{u}_n^{l+1,m} -
  \tilde{b}_n^{l,m}\right] e^{l,m}_n\\
  & + \left\{-p_- e^{l-1,m}_n + \left(1+p_0\right)e^{l,m}_n - p_+ e^{l+1,m}_n -  e^{l,m+1}_n - \lambda\Delta t \cdot
\left(\tilde{I}e^{l,m}_{n-1}\right)^{l,m}
  \right\}
  \left[\tilde{u}^{l,m}- g^l\right]=0.
 \end{split}
 \end{equation}
We can drop the first term on the left-hand-side of (\ref{eq:u -
 u_n}) because of the first inequality in (\ref{eq:diffeq-un}) and
 $e^{l,m}_n$ being nonnegative. It gives us the inequality
 \begin{equation} \label{eq:e_n-ineq-1}
  (1+p_0) e^{l,m}_n \left[\tilde{u}^{l,m}-g^l\right] \leq
  \left[p_- e^{l-1,m}_n + p_+ e^{l+1,m}_n + e^{l,m+1}_n + \lambda\Delta t
  E_{n-1}\right]\left[\tilde{u}^{l,m}-g^l\right],
 \end{equation}
 in which we also used the assumption
 (\ref{eq:ass-rho}) to derive the upper bound for the integral
 term.

 If there are some $(l,m)$ such that $\tilde{u}^{l,m} = g^l$, since
 $\{\tilde{u}_n\}_{n\geq 0}$ is an increasing sequence from Proposition
 \ref{prop:u-n-increase}, we have $\tilde{u}^{l,m} =
 \tilde{u}_n^{l,m}$ for all $n$. Therefore, $e^{l,m}_n = 0$ for
 these $(l,m)$. On the other hand, if $\tilde{u}^{l,m} > g^l$ for
 some $(l,m)$, we can divide $\tilde{u}^{l,m} -g^l$ on both sides
 of (\ref{eq:e_n-ineq-1}) to get
 \begin{equation}\label{eq:e_n-ineq-2}
 \begin{split}
  (1+p_0)e^{l,m}_n & \leq p_- e^{l-1,m}_n + p_+ e^{l+1,m}_n +
  e^{l,m+1}_n + \lambda \Delta t E_{n-1} \\
  & \leq p_- E^m_n + p^+ E^m_n + E^{m+1}_n + \lambda \Delta t
  E_{n-1}.
 \end{split}
 \end{equation}
 Since the right-hand-side of (\ref{eq:e_n-ineq-2}) does not
 depend on $l$, we can write
 \begin{equation}\label{eq:e_n-ineq-3}
  E^m_n \leq \eta E^{m+1}_n +
  (1-\eta)\frac{\lambda}{\lambda+r} E_{n-1},
 \end{equation}
 in which $\eta = \frac{1}{1+(\lambda+r)\Delta t} \in (0,1)$.
 Note that (\ref{eq:e_n-ineq-3}) is also satisfied for all $m$,
 because even if $\tilde{u}^{l,m}=g^l$ for some $(l,m)$, $e^{l,m}_n
 =0$ as we proved above. If follows from (\ref{eq:e_n-ineq-3})
 that
 \begin{equation}\label{eq:e_n-ineq-4}
  E_n^m \leq \eta^{M-m} E^M_m + (1-\eta)(1+\eta + \cdots +
  \eta^{M-m-1}) \frac{\lambda}{\lambda+r} E_{n-1}.
 \end{equation}
 Since the terminal condition of $\tilde{u}_n$, we have
 $E^M_n=0$. Now maximizing the right-hand-side of (\ref{eq:e_n-ineq-4}) over $m$, we obtain that
 \begin{equation*}
  E_n \leq (1-\eta^M) \frac{\lambda}{\lambda+r} E_{n-1}.
 \end{equation*}
 As a result,
 \begin{equation}\label{eq:error-n}
  E_n \leq \left(1-\eta^M\right)^n
  \left(\frac{\lambda}{\lambda+r}\right)^n E_0 \rightarrow 0,
  \quad \text{as } n\rightarrow +\infty.
 \end{equation}
\end{proof}

\begin{remark}\label{remark:dis-conv}
 As $M\rightarrow +\infty$
 \[
  1- \eta^M = 1- \left(\frac{1}{1+(\lambda+r)T/M}\right)^M
  \rightarrow 1- e^{-(r+\lambda)T},
 \]
 which agree with the convergent rate (\ref{eq:iterror}) in the continuous case.
\end{remark}

\begin{proposition}\label{prop:u-tildeu}
 \begin{equation}\label{eq:lim-theta}
  \left|u(x_k, m\Delta t) - \tilde{u}(x_k. m\Delta t)\right|
  \rightarrow 0,
 \end{equation}
 as $\Delta x$, $\Delta t$, $\Delta z \rightarrow 0$.
\end{proposition}
\begin{proof}
 Using the triangle inequality, let us write
 \begin{equation}\label{eq:u-tildeu}
 \begin{split}
  \left|u(x_k, m\Delta t) - \tilde{u}(x_k, m\Delta t)\right|
  \leq & \left|u(x_k, m\Delta t) - u_n (x_k, m\Delta t)\right| +
  \left|u_n(x_k, m\Delta t) - \tilde{u}_n(x_k, m\Delta t)\right| \\
  & +\left|\tilde{u}_n(x_k, m\Delta t) - \tilde{u}(x_k,m\Delta
t)\right| \\
  \leq & K\left(1-e^{-(r+\lambda)(T-m\Delta t)}\right)^n
  \left(\frac{\lambda}{\lambda+r}\right)^n + n\cdot O\left((\Delta t) + (\Delta x)^2 + (\Delta
  z)^2\right) \\
  & + \tilde{K} \left(1-\eta^M\right)^n
  \left(\frac{\lambda}{\lambda+r}\right)^n,
 \end{split}
 \end{equation}
 for some positive constants $K$ and $\tilde{K}$. The first and
 third terms on the right-hand-side of the second inequality are due to (\ref{eq:iterror}) and (\ref{eq:error-conv}). The
 second term arises since the order of error from discretizing a
 PDE using implicit Euler scheme is $O((\Delta t) + (\Delta
 x)^2)$, the interpolation and discretization error from numerical
 integral are of order $(\Delta x)^2$ and $(\Delta z)^2$ and the
 total error made at each step propagates at most linearly in $n$
 when we sequentially discretize (\ref{eq:varineq-un}).

 Letting $\Delta t$, $\Delta x$, $\Delta z\rightarrow 0$ in
 (\ref{eq:u-tildeu}), we obtain that
 \[
  \lim_{\Delta t, \Delta x, \Delta z\rightarrow 0} \left|u(x_k,m\Delta t) - \tilde{u}(x_k, m\Delta
  t)\right| \leq \left(K+\tilde{K}\right) \left(\frac{\lambda}{\lambda +
  r}\right)^n \left(1-e^{-(r+\lambda)T}\right)^n,
 \]
 in which we used (\ref{eq:lim-theta}). Since $n$ is arbitrary the
 result follows.
\end{proof}

\begin{remark}\label{remark:crank-nicolson}
 In Propositions \ref{prop:u-n-increase} - \ref{prop:u-tildeu}, we
 have shown the convergence of the algorithm for completely
 implicit Euler scheme ($\theta=1$). In order to have the time
 discretization error as $O((\Delta t)^2)$, we will choose
 Crank-Nicolson scheme with $\theta=1/2$ in the numerical
 experiments in the next section. From numerical results in Table 4, we shall see that Crank-Nicolson Scheme is also stable and
 the convergence is fast.
\end{remark}

\section{The Numerical Performance of the Proposed Numerical Algorithm}
\label{sec:numericalper}

In this section, we present the numerical performance of the
 algorithm proposed in the previous section. First, we compare the prices we obtain to
the prices obtained in the literature. To demonstrate our
competitiveness we also list the time it takes to obtain the
prices for certain accuracy. We will use either the PSOR or the Brennan-Schwartz algorithm to solve the sparse linear system in (\ref{eq:diffeq-un}); see Remark~\ref{remark:complexity}. All our computations are performed
with C++ on a Pentium IV, 3.0 GHz machine.

In Table 1, we take the jump distribution $F$ to be the double exponential distribution
\begin{equation}\label{eq:double-exp}
F(dz)=\left(p \eta_1 e^{-\eta_1 z}1_{\{z \geq 0\}}+(1-p) \eta_2
e^{\eta_2 z}1_{\{z<0\}}\right)dz.
\end{equation}
We compare our performance with that of \cite{kou-wang} and \cite{kpw}. \cite{kou-wang} obtain an approximate American option price formula, for by reducing the integro-pde equation $V$ satisfied to a integro-ode following \cite{adesi}.
This approximation is accurate for small and large maturities. Also, they do not provide error bounds, the magnitude of which might depend on the parameters of the problem, therefore one might not be able to use this price approximation without the guidance of another numerical scheme.
 A more accurate numerical scheme using an approximation to the exercise boundary and Laplace transform was later developed by \cite{kpw}. Our performance has the same order of magnitude as theirs. Our method's advantage is that it works for a more general jump distribution and we do not have to assume a double exponential distribution for jumps as \cite{kou-wang} and \cite{kpw} do.

In Table 2 we compute the prices of American and European options
in a Merton jump diffusion model, in which the jump distribution
$F$ is specified to be the Gaussian distribution
\begin{equation}
F(dz)=\frac{1}{\sqrt{2 \pi
\tilde{\sigma}^2}}\exp\left(\frac{-(z-\tilde{\mu})^2}{\tilde{\sigma}^2}\right)dz.
\end{equation}
We list the accuracy and time characteristics of the proposed numerical algorithm
algorithm. We compare our prices to the ones obtained by
\cite{dFL,dFV}. \cite{dFL} used a penalty method to approximate
the American option price, while we analyze the variational
inequalities directly (see (\ref{eq:diffeq-u}) and
(\ref{eq:diffeq-un})). Moreover, our approximating sequence is
monotone (see Proposition \ref{prop:u-n-increase}).

In Table~3, We also list the approximated prices of Barrier
options. We compare the prices we obtain with \cite{metwally}
where a Monte Carlo method is used. We do not list the time it
takes for the alternative algorithms in Tables 2 and 3 either
because they are not listed in the original papers or they take
unreasonably long time.

In Table 4, we list the numerical convergence of the proposed algorithm
algorithm with respect to grid sizes. We choose Crank-Nicolson
scheme with $\theta=1/2$ in (\ref{eq:diffeq-un}) and solve the
sparse linear system by either the Bernnan-Schwartz algorithm or the PSOR.

\begin{remark}\label{remark:complexity}
Here we will analyze the complexity of our algorithm. Let us fix
$\Delta x/\Delta t$ as a constant and choose the
 number of grid point in $x$ to be $N$. For each
 time step, using the FFT to calculate the integral term in (\ref{eq:diffeq-un})
 costs $O(N\log{N})$ computations.
 On the other hand, the Brennan-Schwartz algorithm, which uses the LU decomposition to solve the sparse linear system in (\ref{eq:diffeq-un}) (see \cite{j-l-l} pp. 283),
 needs $2N$ computations for each time step. However, PSOR needs
 $C\cdot N$ computations for each time step to solve (\ref{eq:diffeq-un}) at each time step. Here, $C$ is the number of iterations PSOR requires to converge to a fixed small error tolerance
 $\epsilon$. We will see in the following that PSOR is numerically
 more expensive than the Brennan-Schwartz algorithm.

 For PSOR, the number of iterations $C$ increases with respect to $N$. To see this, we
 start from the tri-diagonal matrix on the left-hand-side of (\ref{eq:diffeq-un})
 \begin{equation*}
 A = \left\{
  \begin{array}{cccc}
    1+\theta p_0  & -\theta p_+ & & \\
    -\theta p_- & 1+\theta p_0 & \ddots & \\
    & \ddots & \ddots & -\theta p_+ \\
    & & -\theta p_- & 1+\theta p_0
  \end{array}
  \right\}.
 \end{equation*}
 For the SOR (without projection), the optimal relaxation parameter $\omega$ is given by (see \cite{young})
 \begin{equation*}
  \omega = \frac{2}{1+\sqrt{1-\rho^2_J}},
 \end{equation*}
 where $\rho_J$ is the spectral radius of the Jacobi iteration matrix $J = D^{-1}(A-D)$ with $D$ as the diagonal matrix of $A$. Since $\rho_J \leq \|J\|_{\infty} = \theta (p_+ + p_-)/(1+\theta p_0)$,  we have
 \begin{equation} \label{eq:omega0}
 \omega \leq \omega_0 = \frac{2}{1+\sqrt{1-\|J\|_{+\infty}^2}}.
 \end{equation}
 We will use $\omega_0$ as the optimal relaxation parameter in our numerical experiments. On the other hand, since the largest eigenvalue $\lambda_{max}$ of the SOR iteration matrix is bounded above by $\omega -1$, using (\ref{eq:coeff-p}) and (\ref{eq:omega0}) we obtain that
\begin{equation}\label{eq:C}
C = \min\{c \geq 0 | (\lambda_{max})^c \leq \epsilon\} = O(\sqrt{N}).
\end{equation}

Since $O(N^{3/2})$ dominates $O(N\log{N})$, the complexity of the Iterated PSOR
algorithm at
each time step will be $O(N^{3/2})$. Therefore, with $O(N)$ time steps, the complexity for
Iterated PSOR algorithm is $O(N^{5/2})$. On the other hand, for the Iterated Brennan-Schwartz algorithm,
since $O(N\log{N})$ dominates $O(N)$, the complexity at each time step
will be $O(N\log{N})$. 
Therefore, the complexity of the Iterated Brennan-Schwartz
algorithm is $O(N^2 \log{N})$ since we have $O(N)$ time steps. 

Please refer
to Tables 1, 2, 3 and 4 for numerical performance of both
algorithms.
\end{remark}

Next, we illustrate the behavior of the sequence of functions
$\{v_n(S,t)\}_{n \geq 0}$ and its limit $V$ in Figures 1, 2 and 3.
All the figures are obtained for an American put option in the
case of the double exponential jump with $K=100$, $S_0=100$,
$T=0.25$, $r=0.05$, $\sigma=0.2$, $\lambda=3$, $p=0.6$,
$\eta_1=25$ and $\eta_2=25$ (the same parameters are used in the 8th
row of Table 1) at a single run.
\begin{remark}\label{rem:lstrm}
\begin{enumerate}

\item  In Figure 1, we show, how $V(S,0)$ depends on the time to
maturity, and that it fits smoothly to the put-pay-off function at
$s(0)$ (the exercise boundary). The $y$-axis is the difference
between the option price and the pay-off function. As the time to
maturity increases, the option price $V(S,0)$ increases while the
exercise boundary $s(0)$ decreases. Even though the stock price
process has jumps,  the option price smoothly fits the pay-off
function at $s(0)$, as in the classical Black-Scholes case without
the jumps.

\item In Figure 2, we illustrate the convergence of the exercise
boundaries $t \rightarrow s_{n}(t)$, $n \geq 1$.  We can see from
the figure that all $s_n(t)$ are convex functions. Also, the
sequence $\{s_n\}_{n\geq 1}$ is a monotone decreasing sequence,
which implies that the continuation region is getting larger, and
that the convergence of the free boundary sequence is fast.

Moreover, we notice that,  when the parameters are chosen such
that (\ref{eq:condition-c}) is satisfied, the free boundaries are
discontinuous at the maturity time. In addition, we have
$s(T-)=s_n(T-)=S^*<K$, where $S^*$ is the unique solution of
(\ref{eq:equation-s*}). Furthermore, if $F$ is the double
exponential distribution as in (\ref{eq:double-exp}), the integral
equation (\ref{eq:equation-s*}) can be solved analytically to
obtain
\begin{equation}\label{eq:soln-s*}
 S^* = \left(\frac{(\eta_1 - 1)r}{\lambda p}\right)^{1/\eta_1}
 \cdot K.
\end{equation}
With the parameters we choose, we get from (\ref{eq:soln-s*}) that
$S^* = 98.39$. It is close to our numerical result as one can see
from Figure 2.

\item In Figure 3, we illustrate the convergence of the sequence
of prices $\{v_{n}(S,0)\}_{n \geq 0}$. Observe that this is a
monotonically increasing sequence and it converges to its limit
$V(S,0)$ very fast.
\end{enumerate}
\end{remark}
\hfill \\
\hfill \\

\textbf{Acknowledgment} We are grateful to the two anonymous
referees for their detailed comments that helped us improve our
paper.

\bibliographystyle{plain}
\bibliography{references}

{\small
\begin{table}[h]\label{tab:kou-wang}
\label{table1} \caption{Comparison between the proposed iterated
jump algorithm with the method in \cite{kou-wang} and \cite{kpw},
where the parameters were chosen as $r=0.05$, $S(0)=100$ and
$p=0.6$. Amin's price is calculated in \cite{kou-wang} using the
enhanced binomial tree method as in \cite{amin}. The accuracy of
Amin's price is up to about a penny. The KPW 5EXP price from
\cite{kpw} is calculated on a Pentium IV, 1.8 GHz, while the
iterated price is calculated on Pentium IV, 3.0GHz, both using C++
implementation. Run times are in seconds. For 
numerical algorithm we propose, the number of grid points in $x$ is chosen as $2^6$ and
$\Delta t = \Delta x$. The option prices from both Iterated
Brennan-Schwartz and Iterated PSOR are the same. Below ``B-S" stands for
the Brennan-Schwartz.}

\begin{small}
\begin{minipage}{\textwidth}
\begin{center}
\begin{tabular}{rrrrrr|r|rrrrr|rrrr}
\hline

\multicolumn{16}{c}{American Put Double Exponential Jump Diffusion
Model}\\

\hline \multicolumn{6}{c|}{Parameter Values}& Amin's &
\multicolumn{2}{c}{KW} & \multicolumn{3}{c|}{KPW 5EXP}&
\multicolumn{4}{c}{Proposed Algorithm}\\

K & T & $\sigma$ & $\lambda$ & $\eta_1$ & $\eta_2$ & Price & Value & Error & Value & Error & Time & Value & Error & B-S Time& PSOR Time\\

\hline

90 & 0.25 & 0.2 & 3 & 25 & 25 & 0.75 & 0.76 & 0.01 & 0.74 & -0.01 & 3.21 & 0.75 & 0 & 0.08 & 0.12\\
90 & 0.25 & 0.2 & 3 & 25 & 50 & 0.65 & 0.66 & 0    & 0.65 & 0     & 3.25 & 0.66 & 0.01 & 0.08   & 0.12\\
90 & 0.25 & 0.2 & 3 & 50 & 25 & 0.68 & 0.69 & 0.01 & 0.68 & 0     & 2.97 & 0.69 & 0.01 & 0.08 & 0.12\\
90 & 0.25 & 0.2 & 3 & 50 & 50 & 0.59 & 0.60 & 0.01 & 0.59 & 0     & 2.89 & 0.59 & 0 & 0.12 & 0.12\\
90 & 0.25 & 0.3 & 3 & 25 & 25 & 1.92 & 1.93 & 0.01 & 1.92 & 0     & 2.40 & 1.93 & 0.01 & 0.09 & 0.13\\
90 & 0.25 & 0.2 & 7 & 25 & 25 & 1.03 & 1.04 & 0.01 & 1.02 & -0.01 & 3.18 & 1.03 & 0 & 0.12 & 0.17\\
90 & 0.25 & 0.3 & 7 & 25 & 25 & 2.19 & 2.20 & 0.01 & 2.18 & -0.01 & 2.97 & 2.20 & 0.01 & 0.12 & 0.20\\

\hline

100 & 0.25 & 0.2 & 3 & 25 & 25 & 3.78 & 3.78 & 0   & 3.77 & -0.01 & 3.08 & 3.78 & 0 & 0.12 & 0.12\\
100 & 0.25 & 0.2 & 3 & 25 & 50 & 3.66 & 3.66 & 0   & 3.65 & -0.01 & 3.29 & 3.66 & 0 & 0.10 & 0.12\\
100 & 0.25 & 0.2 & 3 & 50 & 25 & 3.62 & 3.62 & 0   & 3.62 & 0     & 2.88 & 3.63 & 0.01 & 0.09 & 0.12\\
100 & 0.25 & 0.2 & 3 & 50 & 50 & 3.50 & 3.50 & 0   & 3.50 & 0     & 3.00 & 3.50 & 0 & 0.13 & 0.12\\
100 & 0.25 & 0.3 & 3 & 25 & 25 & 5.63 & 5.62 & -0.01 & 5.63 & 0   & 2.44 & 5.63 & 0 & 0.13 &0.15\\
100 & 0.25 & 0.2 & 7 & 25 & 25 & 4.26 & 4.27 & 0.01 & 4.26 & 0    & 3.48 & 4.27 & 0.01 & 0.17 & 0.17\\
100 & 0.25 & 0.3 & 7 & 25 & 25 & 5.99 & 5.99 & 0    & 5.99 & 0    & 2.95 & 6.00 & 0.01 & 0.17 & 0.18\\

\hline

90 & 1 & 0.2 & 3 & 25 & 25 & 2.91 & 2.96 & 0.05 & 2.90 & -0.01 & 2.43 & 2.92 & -0.01 & 0.63 & 0.78\\
90 & 1 & 0.2 & 3 & 25 & 50 & 2.70 & 2.75 & 0.05 & 2.69 & -0.01 & 2.38 & 2.70 & 0 & 0.69 & 0.81\\
90 & 1 & 0.2 & 3 & 50 & 25 & 2.66 & 2.72 & 0.06 & 2.67 & 0.01  & 2.55 & 2.68 & 0.02 & 0.64 & 0.82\\
90 & 1 & 0.2 & 3 & 50 & 50 & 2.46 & 2.51 & 0.05 & 2.45 & -0.01 & 2.30 & 2.45 & -0.01 & 0.68 & 0.82\\
90 & 1 & 0.3 & 3 & 25 & 25 & 5.79 & 5.85 & 0.06 & 5.79 & 0     & 2.48 & 5.77 & -0.02 & 0.70 & 0.94\\

\hline
\end{tabular}
\end{center}
\end{minipage}
\end{small}
\end{table}

\begin{table}[h]
\caption{Option price in Merton jump-diffusion model}
\label{table2}

\begin{minipage}{\textwidth}
\noindent K=100, T=0.25, r=0.05, $\sigma=0.15$, $\lambda=0.1$.
Stock price has lognormal jump distribution with
$\tilde{\mu}=-0.9$ and $\tilde{\sigma}=0.45$. For the iterated
jump schemes, the number of grid points in $x$ is chosen as $2^7$
and $\Delta t = \Delta x$. Below ``B-S" stands for the Brennan-Schwartz.
\begin{center}
\begin{tabular}{r|r|r|rrrr}

\hline

Option Type \footnote{The option prices (for the same kind of option) for different
$S(0)$ are obtained from a single run.} & S(0) & dFLV\footnote{The dFLV price comes from
\cite{dFL,dFV}.} & \multicolumn{4}{c}{Proposed Algorithm} \\
& & & Value & Error & LU(B-S) Time & PSOR Time \\

\hline

American Put & 90 & 10.004 & 10.004\footnote{the option price is 10.001 using the iterated Brennan-Schwartz scheme.} & 0 & 0.18 & 0.24 \\
             & 100 & 3.241 & 3.242  & 0.001 &  \\
             & 110 & 1.420 & 1.420  & 0 &  \\

\hline

European Put & 100 & 3.149 & 3.150 & 0.001 & 0.21 & 0.18 \\

\hline

 European Call & 90 & 0.528 & 0.528 & 0 & 0.18 & 0.18 \\
              & 100 & 4.391 & 4.392 & 0.001 & \\
              & 110 & 12.643 & 12.643 & 0 & \\

\hline
\end{tabular}
\end{center}
\end{minipage}
\end{table}

\begin{table}[t]
\caption{European down-and-out barrier call option
with Merton jump-diffusion model}
\label{table3}

\begin{minipage}{\textwidth}
K=110, S(0)=100, T=1, r=0.05, $\sigma=0.25$, $\lambda=2$, rebate
R=1, the Stock price has lognormal jump distribution with
$\tilde{\mu}=0$ and $\tilde{\sigma}=0.1$. For the algorithm we propose the number of grid points in $x$ is chosen as $2^6$ and
$\Delta t =\Delta x$. Below we use the acronyms LU or SOR to tell wheher we use the LU factorization or the SOR to solve for the  sparse linear systems at each time step. 
\begin{center}
\begin{tabular}{r|r|rrrr}

\hline

Barrier H & MA Price \footnote{The MA price comes from
\cite{metwally}}& \multicolumn{4}{c}{Proposed Algorithm
}\\
& & Value & Error & LU Time & SOR Time\\

\hline

85 & 9.013 & 8.988 & -0.025 & 0.52 & 0.71\\
95 & 5.303 & 5.290 & -0.013 & 0.64 & 0.86\\

\hline
\end{tabular}
\end{center}
\end{minipage}
\end{table}

\begin{table}[h]
\caption{Convergence of the numerical algorithm with
respect to grid sizes}
\label{table4}
\begin{minipage}{\textwidth}
K=100, T=0.25, r=0.05, $\sigma=0.15$, $\lambda=0.1$, stock price
has lognormal jump distribution with $\tilde{\mu}=-0.9$ and
$\tilde{\sigma}=0.45$ (the same parameters that are used in
\cite{dFL}). The differential equation is discretized by the
Crank-Nicolson scheme as (\ref{eq:diffeq-un}) with $\theta =1/2$.
The logarithmic variable $x=\log S$ is equally spaced discretized
on an interval $[x_{min}, x_{max}]$ with $\Delta x = \Delta t$.
The numerical integral is truncated on the smallest interval
$[z_{min}, z_{max}]$, such that $[x+\tilde{\mu}-4\tilde{\sigma},
x+\tilde{u}+4\tilde{\sigma}]$ will be inside $[z_{min}, z_{max}]$
for any $x\in [x_{min}, x_{max}]$.  The step length for the
numerical integral is chosen the same as the step length in $x$,
i.e. $\Delta z =\Delta x$. The number of grid points for to
implement the FFT is chosen as an integral power of 2. The error
tolerance for PSOR method is $10^{-8}$ and for the global
iteration is $10^{-6}$. Run times are in seconds. Each row in the
``Difference" column of the following table is $v_{PSOR}(L,M) -
v_{PSOR}(L/2,M/2)$. ``B-S" stands for the Brennan-Schwartz
algorithm. The number of global iteration is 3 for all the
following numerical experiments.
\begin{small}

\begin{center}
\begin{tabular}{|c|c|c|c|c|c|c|c|c|}

\hline

S(0) & No. of grid   & No. of time & B-S Value & B-S & PSOR Value & Difference & PSOR  &  Max. No. of \\
 & points in $x$ ( L ) & steps ( M ) & $v_{B-S}$ & Time & $v_{PSOR}$ & & Time &   PSOR iterations\\
 \hline

 \multirow{4}{*}{90} & 64 & 30 & 10.00230 & 0.06 & 10.00573 & n.a. & 0.06 & 16 \\
 & 128 & 58 & 10.00142 & 0.21 & 10.00429 & -0.00144 & 0.24 &  21 \\
 & 256 & 115 & 10.00192 & 0.84 & 10.00396 & -0.00033 & 0.99 &  28 \\
 & 512 & 230 & 10.00218 & 3.51 & 10.00387 & -0.00009 &  4.50 &  39\\
 \hline

 \multirow{4}{*}{100} & 64 & 30 & 3.24074 & 0.06 & 3.24465 & n.a. & 0.06 &  16 \\
 & 128 & 58 & 3.24008 & 0.21 & 3.24180 & -0.00285 & 0.24 &  21 \\
 & 256 & 115 & 3.24046 & 0.84 & 3.24115 & -0.00065 & 0.99 &  28\\
 & 512 & 230 & 3.24058 & 3.51 & 3.24103 & -0.00012 & 4.50 &  39\\
 \hline

 \multirow{4}{*}{110} & 64 & 30 & 1.42048 & 0.06 & 1.42146 & n.a. & 0.06 &  16\\
 & 128 & 58 & 1.41941 & 0.21 & 1.41991 & -0.00155 & 0.24 &  21\\
 & 256 & 115 & 1.41958 & 0.84 & 1.41966 & -0.00025 & 0.99 &  28\\
 & 512 & 230 & 1.41962 & 3.51 & 1.41960 & -0.00006 & 4.50 &  39\\
 \hline
\end{tabular}
\end{center}

Using (\ref{eq:C}), the number of SOR iterations can be
calculated. The calculation gives $11, 16, 22 $ and $31$.
Comparing with the last column of above table, the maximum numbers
of PSOR iteration are slightly larger than these theoretical
predicted SOR iteration times. Moreover, when $L = 512$ the the
ratio between the maximum number of PSOR iteration and $\sqrt{L}$
is 1.72. This confirms the analysis in Remark
\ref{remark:complexity} that the maximal PSOR iteration time
grows as the order of $\sqrt{L}$.

\end{small}
\end{minipage}
\end{table}

}

\begin{figure}[h]
The parameters for the following three figures are $K=100$,
$S_0=100$, $T=0.25$, $r=0.05$, $\sigma=0.2$, $\lambda=3$, the
stock price has double exponential jump with $p=0.6$, $\eta_1=25$
and $\eta_2=25$ (the same parameters used in the 8th row of Table
1).
\caption{ The option price function $S \rightarrow V(S,0)$
smoothly fits the pay-off function $(K-S)^+$ at $s(0)$. $V(S,0)$
increases and $s(0)$ ($V(S,0)-(K-S)^+=0$ at s(0)) decreases as
time to maturity $T$ increases.}
\includegraphics[width=6.25in, height=3.5in]{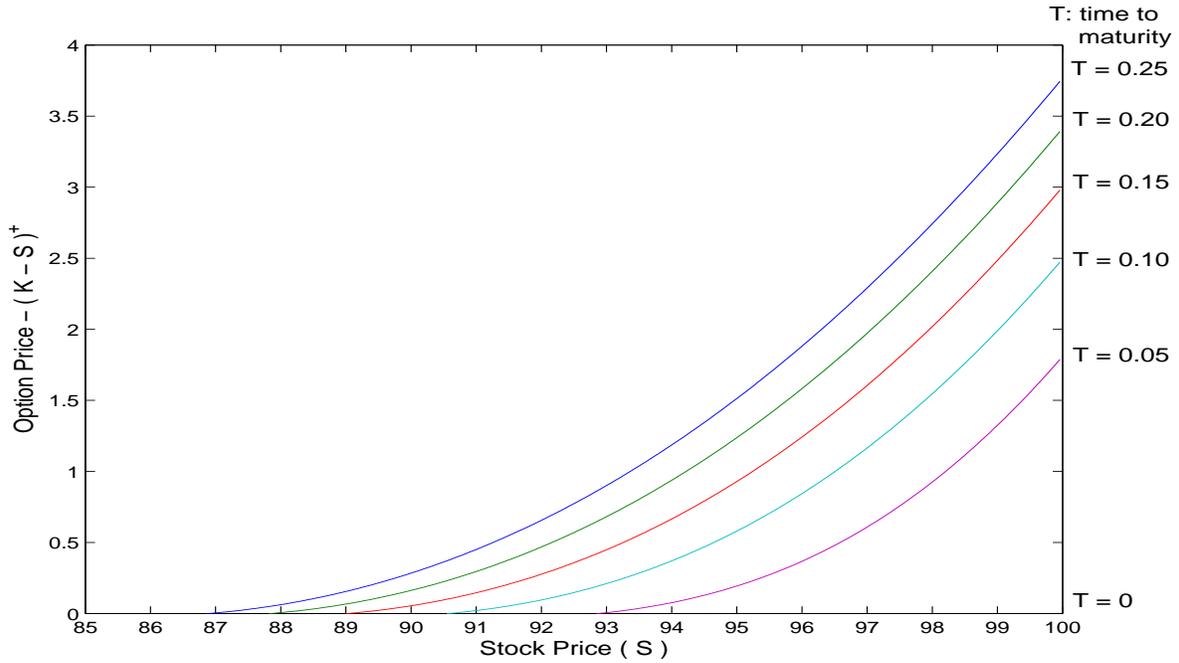}

\end{figure}

\begin{figure}[h]

\caption{Iteration of the Exercise Boundary: $s_n(t) \downarrow
s(t)$, $t \in [0,T)$. Both $s_n(t)$ and $s(t)$ will converge to
$S^* < K$ as $t\rightarrow T$.}
\includegraphics[width=6.5in, height=3.5in]{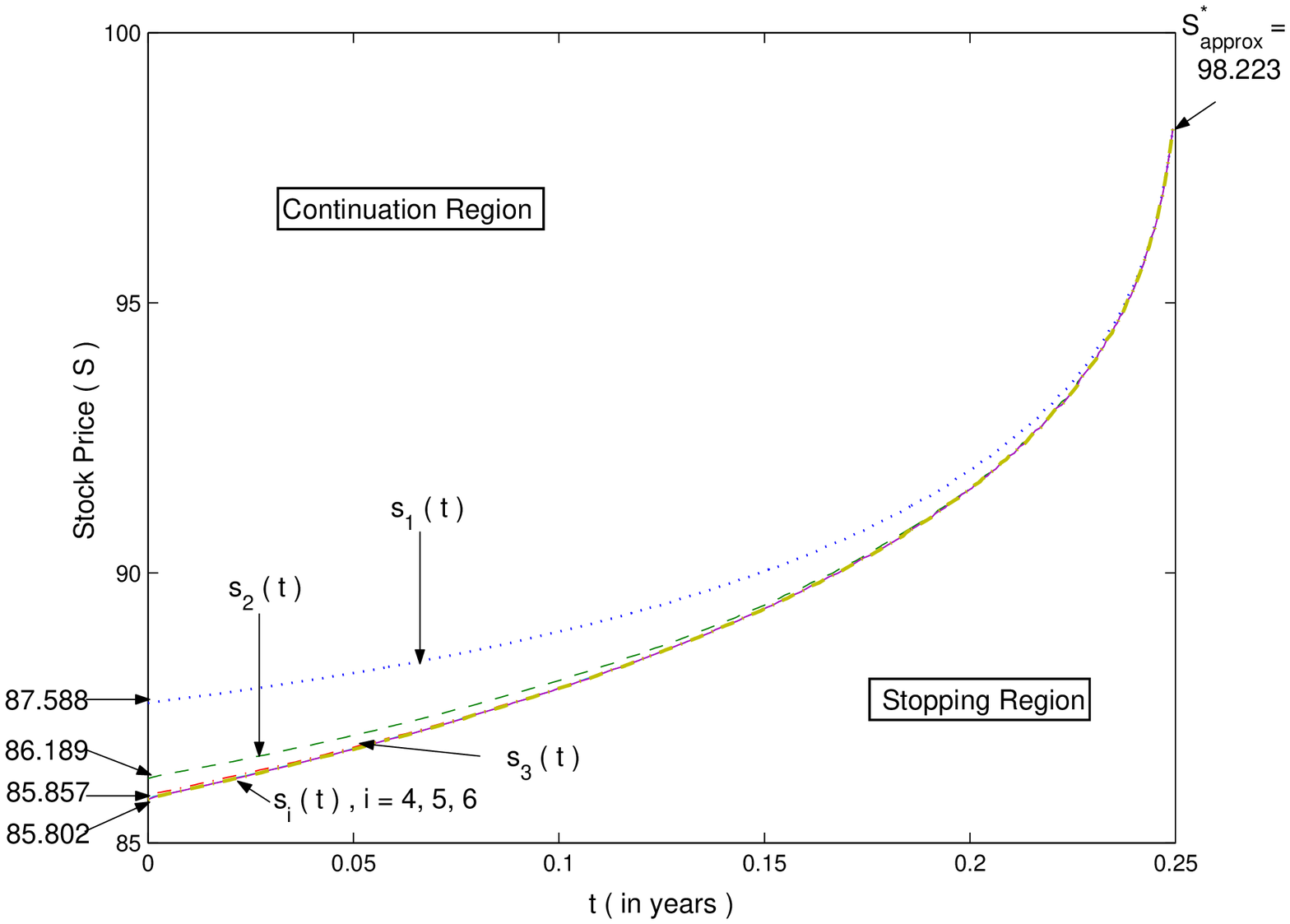}

\caption{Iteration of the price functions: $v_{n}(S,0) \uparrow
V(S,0)$, $S \geq 0$.}

\includegraphics[width=6in, height=3.5in]{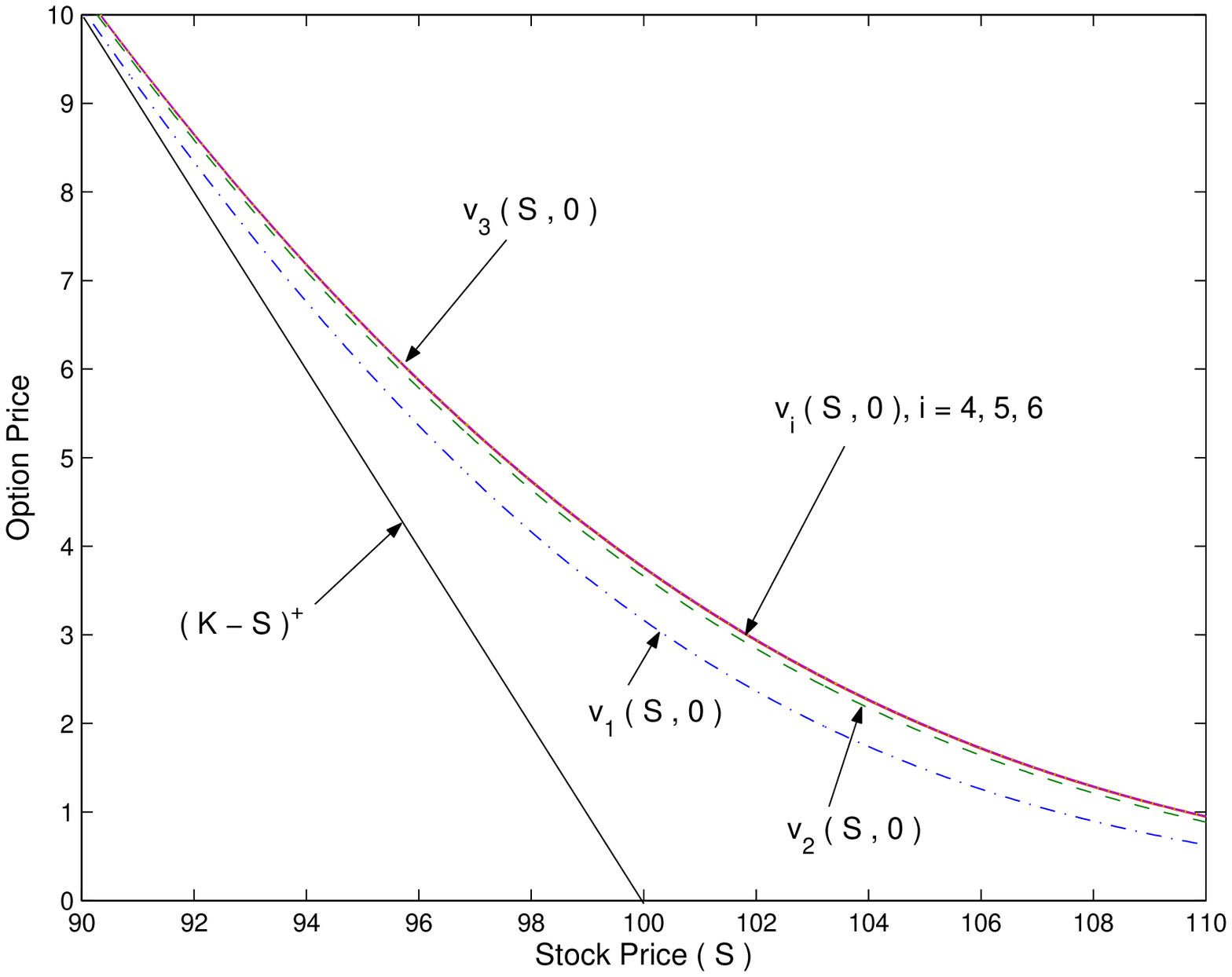}

\end{figure}

\end{document}